\documentclass[sn-mathphys]{sn-jnl}% Math and Physical Sciences Reference Style
\usepackage{hyperref}
\usepackage{url}
\jyear{2021}%

\theoremstyle{thmstyleone}%
\theoremstyle{thmstyletwo}%

\theoremstyle{thmstylethree}%

\begin{document}

\title[Lab-scale Vibration Analysis Dataset]{Lab-scale Vibration Analysis Dataset and Baseline Methods for Machinery Fault Diagnosis with Machine Learning}

%%=============================================================%%
%% Prefix	-> \pfx{Dr}
%% GivenName	-> \fnm{Joergen W.}
%% Particle	-> \spfx{van der} -> surname prefix
%% FamilyName	-> \sur{Ploeg}
%% Suffix	-> \sfx{IV}
%% NatureName	-> \tanm{Poet Laureate} -> Title after name
%% Degrees	-> \dgr{MSc, PhD}
%% \author*[1,2]{\pfx{Dr} \fnm{Joergen W.} \spfx{van der} \sur{Ploeg} \sfx{IV} \tanm{Poet Laureate}
%%                 \dgr{MSc, PhD}}\email{iauthor@gmail.com}
%%=============================================================%%

\author*[1,2]{\fnm{Bagus Tris} \sur{Atmaja}}\email{b-atmaja@aist.go.jp}

\author[1]{\fnm{Haris} \sur{Ihsannur}}\email{harisihsannur@gmail.com}
% \equalcont{These authors contributed equally to this work.}

\author[1]{\fnm{Suyanto} \sur{}}\email{suyanto@ep.its.ac.id}
% \equalcont{These authors contributed equally to this work.}

\author[1]{\fnm{Dhany Arifianto} \sur{}}\email{dhany@ep.its.ac.id}

\affil*[1]{\orgdiv{Department of Engineering Physics}, \orgname{Sepuluh Nopember Institute of Technology}, \orgaddress{\street{ITS Sukolilo Campus}, \city{Surabaya}, \postcode{60111}, \state{Jawa Timur}, \country{Indonesia}}}

\affil[2]{\orgname{National Institute of Advanced Industrial Science and Technology}, \city{Tsukuba}, \postcode{3058560}, \country{Japan}}

% \affil[3]{\orgdiv{Department}, \orgname{Organization}, \orgaddress{\street{Street}, \city{City}, \postcode{610101}, \state{State}, \country{Country}}}

%%==================================%%
%% sample for unstructured abstract %%
%%==================================%%

\abstract{The monitoring of machine conditions in a plant is crucial for production in manufacturing. A sudden failure of a machine can stop production and cause a loss of revenue. The vibration signal of a machine is a good indicator of its condition. This paper presents a dataset of vibration signals from a lab-scale machine. The dataset contains four different types of machine conditions: normal, unbalance, misalignment, and bearing fault. Three machine learning methods (SVM, KNN, and GNB) evaluated the dataset, and a perfect result was obtained by one of the methods on a 1-fold test. The performance of the algorithms is evaluated using weighted accuracy (WA) since the data is balanced. The results show that the best-performing algorithm is the SVM with a WA of 99.75\% on the 5-fold cross-validations. The dataset is provided in the form of CSV files in an open and free repository at https://zenodo.org/record/7006575.}

\keywords{Vibration data, vibration analysis, predictive maintenance, machine condition monitoring, anomaly detection, machine learning}

%%\pacs[JEL Classification]{D8, H51}

%%\pacs[MSC Classification]{35A01, 65L10, 65L12, 65L20, 65L70}

\maketitle

\section{Introduction}\label{sec1}
Vibration analysis is the process of evaluating the vibration characteristics of a machine or structure, typically with the goal of identifying any problems or abnormalities that may be present. Vibrations are often indicative of the health and performance of a machine or structure and can provide valuable information about the condition of certain components, such as bearings, gears, and motors. By analyzing the characteristics of vibrations, such as frequency, amplitude, and waveform, it is possible to identify potential problems or failures that may occur in the future. The analysis of vibration is often performed in the frequency domain since the pattern of abnormalities in this domain is more obvious than in the time domain.

% other predictive maintenance methods, but focus on vibration analysis
Vibration signals convey more information than others for predictive maintenance, a maintenance technique based on the condition of machines. Other techniques are oil (lubricant) analysis \cite{Girdhar2004}, infrared thermography \cite{Bagavathiappan2013}, and sound pattern analysis \cite{Delgado-Arredondo2017,Glowacz2018,Atmaja2009a}. Vibration and lubricant analysis were the most common techniques for predictive maintenance (PdM) \cite{carnero2007model}. PdM, which is developed in the 1970s, is an advancement of preventive maintenance, a time-based maintenance from the 1950s \cite{Innovation2015}. Vibration analysis is a key predictive maintenance technique (among others) since it can identify the problem of machines before they become too serious and cause unscheduled downtime \cite{Girdhar2004}.

% the need of lab-scale data
Current technologies in vibration analysis lack in many aspects. The requirement for a large amount of data is challenging \cite{Ypma2001} and can be difficult for real machines (for treating machines in different conditions). The accuracy and reliability of vibration analysis depend on the quality of the sensors and measurement equipment being used. Vibration analysis requires specialized knowledge and expertise to interpret the data and identify potential problems \cite{Ypma2001}. Finally, vibration analysis can be expensive due to the specialized equipment and software required. This paper is presented to tackle the limitations of vibration analysis above.

% since the pattern is obvious, let use ML instead of DL
The use of machine learning to replace expert engineers has been tried previously, including deep learning methods. In \cite{Ypma2001}, the authors used self-organizing maps (SOM) for a leak detection problem based on vibration signals. A jump to using deep learning for vibration-based machinery fault detection has been tempted in \cite{Yang2019,Zhao2019a, Neupane2021}. In \cite{Yang2019}, the authors proposed a deep transfer learning based on vibration data converted to images from VGG-16 to their vibration data. In \cite{Zhao2019a}, the authors reviewed machine deep learning methods for vibration analysis: auto-encoder (AE), deep belief network (DBN), deep Boltzmann machines (DBM), convolutional neural network (CNN), and recurrent neural network (RNN). In \cite{Neupane2021}, the authors proposed a switchable normalization CNN for bearing fault detection. In this light, we saw a gap between previous old SOM methods and recent deep learning methods.

% why machine learning
Since the vibration patterns (the extracted feature) of each machine condition are distinct \cite{Ypma2001}, it is arguably better and enough to use machine learning instead of deep learning. Machine learning seeks to find relatively small data patterns given the features instead of patterns in the data itself, as in deep learning (which also was usually used to extract the features). If the extracted features from vibration signals are informative enough to distinguish the machine conditions, machine learning can be used to detect the machine conditions. In this paper, we evaluated three machine learning, namely support vector machine (SVM), K-nearest neighbors (KNN), and Gaussian Naive Bayes (GNB), to detect the machine conditions-based features extracted from vibration signals.

Instead of evaluating machine learning only, the need for a free dataset for vibration analysis is mainly addressed. The dataset is collected from a lab-scale vibration analysis experiment. Lab-scale data is often used in vibration analysis to validate and verify the results of simulations and calculations, as well as to confirm the performance of new designs or technologies. Lab-scale testing allows researchers and engineers to study the behavior of a system under controlled conditions, which can provide valuable insights into the problem and its potential solution for the system. Lab-scale testing can also be used to assess the feasibility and reliability of a design or technology before it is implemented on real data. This lab-scale data can help and allow designers and engineers to make necessary adjustments or improvements for the development of vibration analysis tools.

This paper, hence, contributes in two aspects. First, we provided a free vibration dataset in CSV format that can be downloaded directly from the open repository. Second, we provided baseline methods with machine learning to detect the machine conditions based on the vibration signals. Furthermore, we show that our evaluation of the dataset and methods achieve a near-perfect accuracy on 5-fold cross-validation and a perfect accuracy on 1-fold test data based on a set of distinct features, highlighting the effectiveness of the proposed dataset and methods.

\section{Previous Works: The Available Datasets}
Research on vibration analysis has been conducted progressively over the years. However, the lack of reports on the available datasets for vibration analysis is a challenge. The available datasets are not well documented, and some datasets are not available for free. The focus of this section is to review the datasets for vibration analysis based on the literature.

MaFaulDa (Machinery Fault Diagnosis) is the openly available vibration analysis dataset from the Signals, Multimedia, and Telecommunications Laboratory, Universidade Federal do Rio de Janeiro. The dataset contains 1951 samples from six conditions: normal, horizontal misalignment, vertical misalignment, imbalance, underhang bearing, and overhang bearing. The apparatus for data collection is  SpectraQuest's Machinery Fault Simulator (MFS) Alignment-Balance-Vibration (ABVT) system. The data is collected at a sample rate of 51.2 kHz from two IMI sensors (model 601A01 and 604B31). In addition to vibration data, the dataset also provides tachometer signals (to estimate rotation frequency) and microphone data (Shure SM81). This dataset was used in several pieces of research, such as in \cite{Sokolovsky2021,Nath2021,Marins2018,Ribeiro2017}. 

Case Western Reserve University (CWRU) \cite{Ribeiro} bearing fault dataset is a dataset designed for examining normal and faulty conditions of the ball bearing. The experiments to obtain the dataset were conducted using a two horsepower (hp) Reliance Electric motor, and acceleration data was measured at locations near to and remote from the motor bearings \cite{CWRU_bearing}. The faults on the motor bearing were created using electro-discharge machining (EDM). The CWRU dataset contains bearing faults ranging from 0.007 inches in diameter to 0.040 inches in diameter and was introduced separately at the inner raceway, rolling element (i.e., ball), and outer raceway. The vibration data was recorded for motor loads, in which faulted bearings were installed, of 0 to 3 horsepower (motor speeds of 1797 to 1720 RPM). This dataset were used in \cite{Marins2018,Neupane2021,Ribeiro2017,Toh2020}.

Another toy dataset is Accelerometer Dataset, which is developed by Mackenzie Presbyterian University, Sao Paulo, Brazil, for predictions of the failure time of a cooling fan \cite{Sampaio2019}. As an apparatus is a cooling fan (Akasa AK-FN059) with weights on its blades was used to generate vibrations. This fan cooler was attached an accelerometer, MMA8452Q accelerometer, to collect the vibration data. There are 153000 records (lines) of vibration data in this dataset.
 
It is interesting to see that each dataset has a different goal. The MaFaulDa dataset is the closest to our dataset for machine condition diagnosis, with differences in apparatus and number of data. Here, we used a lab-scale vibration analysis apparatus (real electrical motors) to collect the vibration data. The vibration data is collected from five machines with different conditions. We designed the dataset to fill the gap in the available datasets for vibration analysis. Another dataset, such as CWRU dataset, focuses on bearing only, while Accelerometer Dataset focuses on predicting the failure time of a cooling fan.

\section{Dataset: VBL-VA001}
\subsection{Recorded conditions}
The need for a Lab-scale vibration analysis dataset is triggered by the difficulties of obtaining the actual data from a real plant. Treating a machine to fail will cost a lot of money and time. Using lab-scale data will minimize the impact of interfering with machine operation while keeping the same fault pattern as the real plant data. In this light, we simulated four common conditions of machine operation by electric motors (water pumps) to collect their vibration patterns.

\begin{figure}[htbp]
    \centering
    \includegraphics[width=\textwidth]{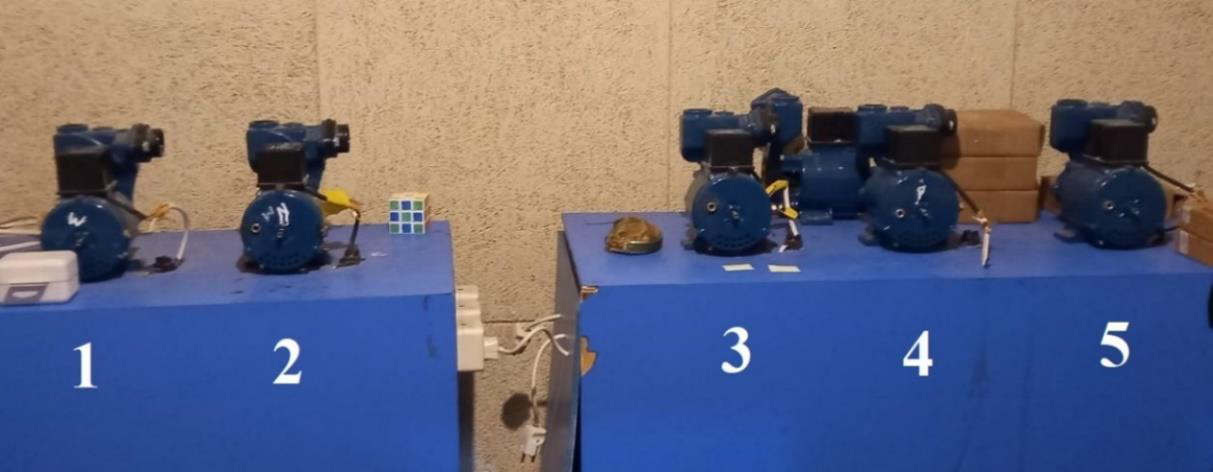}
    \caption{Five electrical machines with different fault conditions: (1) Misalignment, (2) Normal, (3) Unbalance 27 gram.cm, (4) Bearing fault, and (5) Unbalance 6 gram.cm.}
    \label{fig:pump}
\end{figure}

Five electrical motors were designed to replicate four machine conditions: a machine for a condition except for unbalance, where two machines simulate different weights of unbalance conditions. The configuration of these five machines is shown in Figure \ref{fig:pump}. The first machine is with misalignment, the second is with the normal condition, the third is with 27 gram.cm unbalance, the fourth is with bearing fault, and the fifth is with 6 gram.cm unbalance.

\begin{figure}[htbp]
    \centering
    \includegraphics[width=\textwidth]{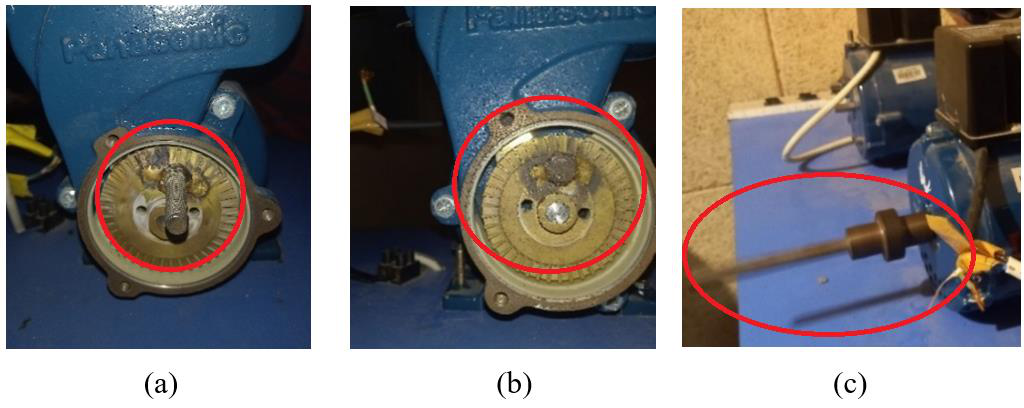}
    \caption{Condition for unbalance and misalignment: (a) adding 18 grams of mass, (b) adding 4 grams of mass, (c) coupling shaft with an additional metal cylinder.}
    \label{fig:pump-fault}
\end{figure}

Unbalance condition is given in two different mass additions to the impeller. The impeller has a diameter of 6 cm. Hence, adding 4 grams of mass from 1.5 of center mass (eccentricity) will cause an unbalance of 6 gram.cm ($ 4 \times 1.5 = 4$). Consequently, adding 18 grams of mass from 1.5 of center mass will cause an unbalance of 27 gram.cm ($ 18 \times 1.5 = 27$). These configurations are retained from the previous work \cite{Taufan2018,Haris2022}. Figs. \ref{fig:pump-fault} (a) and (b) show these unbalance conditions.

The misalignment condition is given by coupling the shaft with an additional metal cylinder. The metal cylinder is 1 cm in diameter and 7 cm in length. The misalignment is 3 mm offset from the center. This misalignment condition was set between the original shaft and an additional metal cylinder with a metal cylinder offset from the shaft. Fig. \ref{fig:pump-fault} (c) shows the misalignment condition.

The bearing fault condition was set by hitting the outer ring of the bearing with a hammer. Hence, the fault is caused by the impact of the hammer (crack-like faults). Although only the outer ring was hit by the hammer, both rings (inner and outer) showed to be in fault conditions simultaneously from the spectrum visualization. It is not possible to detect either inner or outer ring faults only in this dataset since the goal is to detect the general pattern of bearing faults. Future research may be able to detect the specific fault of the bearing, which is one of the most faults in the machine (like dataset in \cite{CWRU_bearing}).

% add the url for download
Table \ref{tab:vbl-va001-data} summarizes our VBL-VA001 dataset. In total, there are 4000 vibration samples, with 1000 samples for each condition. Comparing the existing dataset (Table \ref{tab:data-comparison}), VBL-VA001 is the largest dataset with the most number of samples. The VBL-VA001 dataset is available at \url{https://zenodo.org/record/7006575#.Y5wlTafP2og}. From the original IDE format, we provided our data in CSV format for convenience.

% need clarification for bearing fault: outer only or both

\begin{table}
    \caption{Data distribution of VBL-VA001.}
    \centering
    \begin{tabular}{l l c }
        \hline
        \multicolumn{2}{l}{ Condiditon } & \# Samples \\
        \hline
        \multicolumn{2}{l}{ Normal } & 1000 \\
        Unbalance & 6 gram.cm & 500 \\
        & 27 gram.cm & 500 \\
        Misalignment & $3.0 \mathrm{~mm}$ & 1000 \\
        Bearing Fault & Outer Ring & 1000 \\
        \hline
        \multicolumn{2}{c}{ Total } & $\mathbf{4 0 0 0}$ \\
        \hline
        \end{tabular}
        \label{tab:vbl-va001-data}
    \end{table}

    \begin{table}[htbp]
        \caption{Comparison of VBL-VA001 with other datasets.}
        \centering\begin{tabular}{l c c c c}
        \hline
        Dataset & \# Samples & Sampling Freq. (kHz) & \# Classes \\
        \hline
        VBL-VA001 & 4000 & 20 & 4 \\
        MaFaulDa & 1951 & 50 & 4 \\
        CWRU & 161 &  12 \& 48 & 2 \\
        \hline
        \end{tabular}
        \label{tab:data-comparison}
    \end{table}

\subsection{Sensor placement}
\begin{figure}[htbp]
    \centering
    \includegraphics[width=\textwidth]{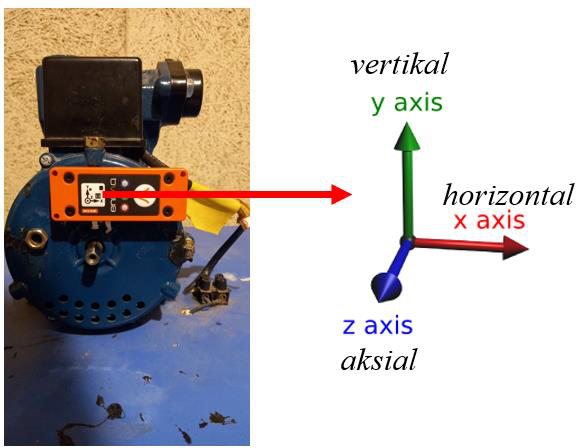}
    \caption{Sensor placement for vibration measurements}
    \label{fig:Sensor}
\end{figure}

The sensor is located in the machine (electrical machine/water pump) in the position as shown in Fig. \ref{fig:Sensor}. The vibration sensor is "LOG-0002-
100G-DC-8GB-PC Shock and Vibration Sensor". To attach it to the machine, we used double-sided tape to mount the sensor into the machine and connected it to the PC with a USB cable. The recording process was done using enDAQ LAB Software. By this arrangement, we collected three axes of acceleration data that comply with the standard (ISO13373-1, 2002). The sensor is located on the rigid part and close to the vibration source. The first consideration (rigid surface) is to avoid resonance, while the second consideration (close to the source) is to minimize the effect of the transmission path. The data were recorded every five seconds. The example of data collected by the sensor is shown in Table \ref{tab:example-data}.

\label{sec:dataset}

% Give an example of data
\begin{table}[htbp]
    \centering
    \caption{Excerpt of data collected by the sensor.}
    \begin{tabular}{c c c c}
    \hline
    Time & X & Y & Z \\
    \hline
    0.004516 & -0.102961 &  0.030537 &  0.114270 \\
    0.004566 & -0.118802 &  -0.020894 &  0.123060 \\
    0.004616 & -0.110881 &  -0.046609 &  0.114270 \\
    0.004666 & -0.102961 &  -0.053038 &  0.105480 \\
    0.004716 & -0.087121 &  -0.059466 &  0.096690 \\
    ... & ... & ... & ... \\
    4.999968 & 0.039601 & -0.083574 & -0.101085 \\
    \hline
    \end{tabular}
    \label{tab:example-data}
\end{table}

\section{Machine Learning Methods}
\label{sec:ml}
\begin{figure}[htbp]
    \centering
    \includegraphics[width=\textwidth]{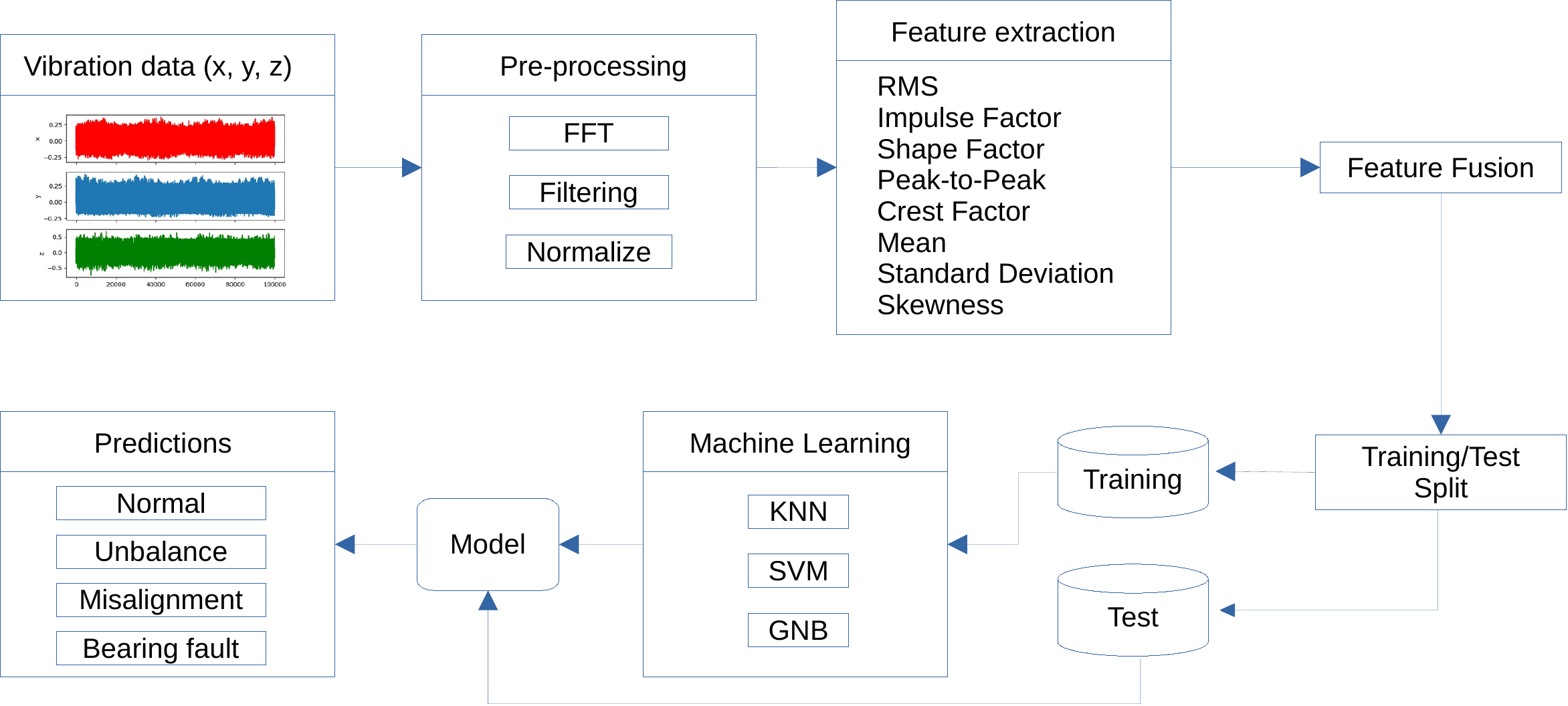}
    \caption{Flowchart of processing the vibration data with machine learning methods; the filtering process in pre-processing remove NaN (not-a-number) values; each feature in the feature extraction process has three values (x, y, z); hence, the total feature (feature fusion) has 27-dim (9 features $\times$ 3 axes).)}
    \label{fig:flowchart}
\end{figure}

The flow of vibration data processing is shown in Fig. \ref{fig:flowchart}. The vibration data is first preprocessed to convert from time-domain signals to frequency-domain signals. We performed data normalization after that since machine learning methods are sensitive to the range of input data. Then, the features are extracted. The extracted features are then used to train the machine-learning model. The trained model is then used to predict the fault condition of the machine in the test data.

\subsection{Pre-Processing and Feature Extractions}
\label{sec:feat}
\subsubsection{FFT}
\begin{figure}[htbp]
    \centering
    \includegraphics[width=\textwidth]{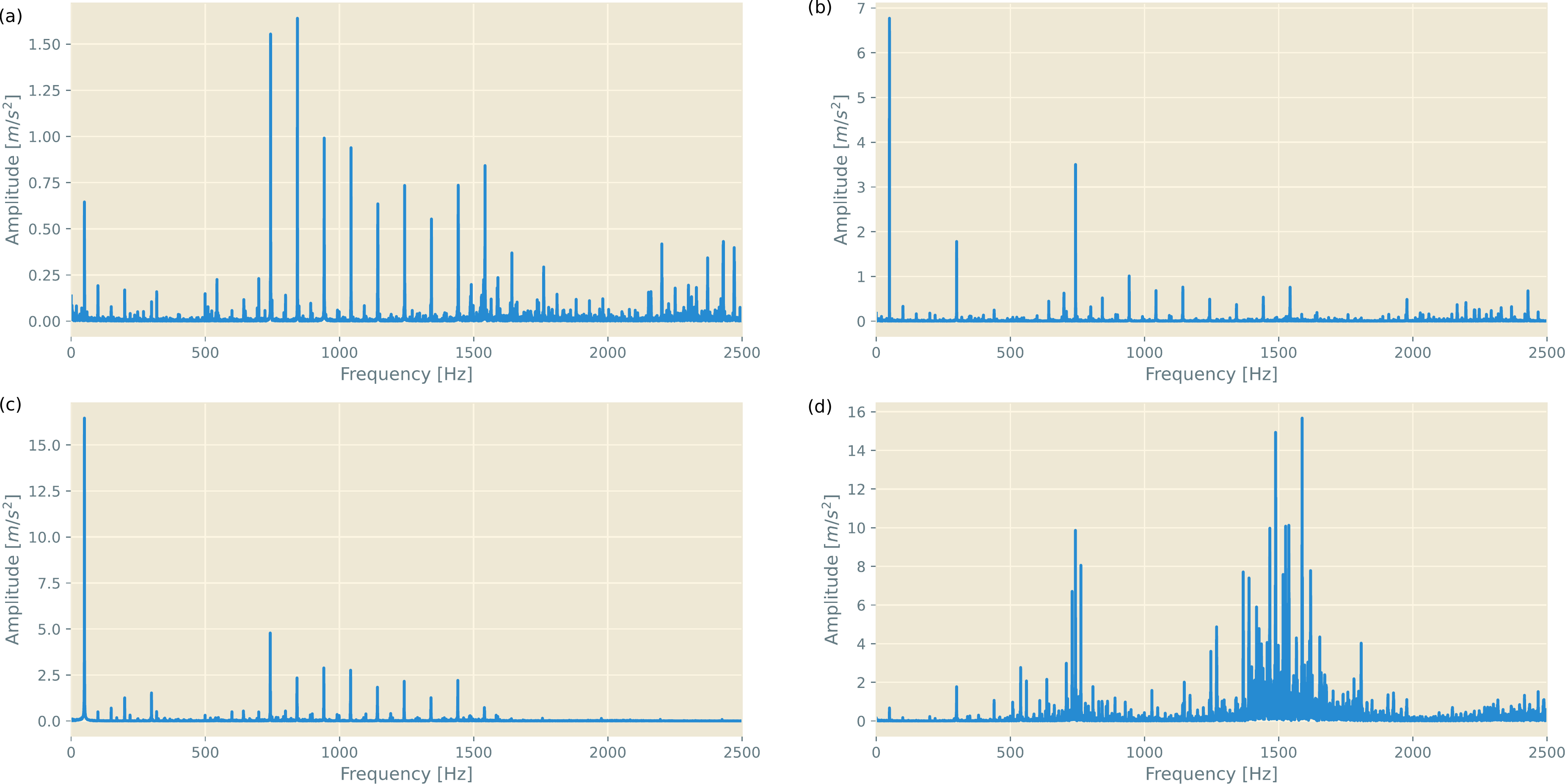}
    \caption{Spectrum of vibration signal in each machine condition: (a) normal, (b) unbalance, (c) misalignment, (d) bearing fault}.
    \label{fig:fft}
\end{figure}

The main pre-processing data are FFT and filtering. FFT is used to convert the time-domain signal into a frequency-domain signal. We used the FFT package from Numpy for this purpose with the default configuration. Figure \ref{fig:fft} shows the results of FFT for each machine condition (shown for the z-axis only). It can be seen that the amplitude of the normal condition is lower than the faulty condition. In that figure, we did not normalize the amplitude of the signal to show the difference between each machine condition. However, the limit of frequency is set to 2500 Hz for clarity of comparison. From the original data provided in acceleration (g) units, we showed the unit in $m/s^2$.

\subsubsection{Normalization}
Since machine learning methods are sensitive to the range of input data, we normalized the data after merging all samples. We used the following
formula to normalize the data:
\begin{equation}
    x_{norm} = \frac{x - x_{min}}{x_{max} - x_{min}}
\end{equation}
where $x_{norm}$ is the normalized data, $x$ is the original data, $x_{min}$ is the minimum value of the data, and $x_{max}$ is the maximum value of the data. The normalized data is then used to extract the features.

\subsubsection{Extracted Features}
The vibration data after the FFT process is simplified using nine feature extraction methods, as shown in Table \ref{tab:feature}. The goal of the feature extraction is for dimensional reduction as well as finding a correlation between the specific patterns of machine conditions. The features are statistics extracted from the frequency domain signal (x, y, z) and then merged into one feature vector. Statistical descriptors are known to be useful for SVM-based machine condition classification \cite{Ebrahimi2017}. In this research, the feature vector has 27 dimensions (9 features $\times$ 3 axes) compared to 44 dimensions in the aforementioned literature.

Finally, we performed filtering by removing outlier data using Pandas' `dropna' method. This step is performed after the feature extraction process and before feeding extracted features to machine learning methods. These outlier data are recorded by the measurements process that returns NaN values. It is not clear why the NaN values are recorded, but it is safe to remove them since they could trigger an error when using the machine learning tool.

\begin{table}
\centering
\caption{Feature extraction methods}
\begin{tabular}{l c}
    \hline Feature & Formula\\
    \hline
    Mean & $\displaystyle \frac{1}{N} \sum_{i=1}^N x_i$ \\
    Standard Deviation (std) & $\displaystyle \sqrt{\frac{1}{N} \sum_{i=1}^N\left(x_i-\bar{x}\right)^2}$ \\
    Root Mean Square (RMS) & $\displaystyle \sqrt{\left(\frac{1}{N}\right) \sum_{i=1}^N(x_i)^2}$ \\
    Peak to Peak (PP) & $\displaystyle x_{max}- x_{min}$ \\
    Impulse Factor (IF) & $\displaystyle \frac{x_{max}}{\bar{x}}$ \\
    Skewness (S) & $\displaystyle \frac{1}{N} \sum_{i=1}^N \frac{\left(x_i-\bar{x}\right)^3}{\sigma^3}$ \\
    Kurtosis (K) & $\displaystyle \frac{1}{N} \sum_{i=1}^N \frac{\left(x_i-\bar{x}\right)^4}{\sigma^4}$ \\
    Crest Factor (C) & $\displaystyle \frac{\left\lvert x_{max} \right\rvert}{RMS}$ \\
    Shape Factor (SF) & $\displaystyle \frac{1}{\frac{1}{N} \sum_{i=1}^N x_i}$ \\
    % Total & & 27\\
    \hline
    \end{tabular}
    \label{tab:feature}
\end{table}

\subsection{Classifiers}
Machine learning is a field of artificial intelligence that involves training computers to perform tasks without explicit programming. It is based on the idea that systems can learn from data, identify patterns, and make decisions with minimal human intervention. At the heart of machine learning is a classifier, the method to classify inputs into outputs. There are a lot of methods (classifiers) developed for machine learning and counting. In this study, we evaluated three machine learning methods, namely support vector machine (SVM), K-nearest neighbors (KNN), and Gaussian naive Bayes (GNB).

\subsubsection{Support Vector Machine}
Support Vector Machines (SVMs) are a type of supervised learning algorithm that can be used for classification (support vector classification, SVC) or regression (support vector regression, SVR). The goal of an SVM is to find the hyperplane in a high-dimensional space that maximally separates the two classes.
An SVM model is trained by finding the hyperplane that has the greatest distance (called the margin) between the two classes. Data points that are closest to the hyperplane are called support vectors and have the greatest influence on the position of the hyperplane. In this study, we used SVC to classify the machine condition.

One of the key strengths of SVMs is their ability to use kernels, which are functions that can transform the data into a higher-dimensional space in which it may be more linearly separable. This allows SVMs to model complex relationships in the data and can lead to improved performance. This study used the radial basis function (RBF) kernel, which is the default kernel in the scikit-learn implementation of SVM (\cite{scikit-learn}). We optimized the regularization parameter $C$ (in a range [0, 100]) with 1-fold and 5-fold cross-validation. The best $C$ value is selected based on the highest accuracy score during the training phase.

\subsubsection{K-nearest neighbors}
K-nearest neighbors (KNN) is a machine learning algorithm that is used for classification and regression. It works by finding the $K$ nearest data points to a given data point and using those points to make a prediction. KNN is a simple and effective algorithm that is easy to implement and works well on a variety of datasets. However, it can be computationally expensive, especially for large datasets, and it can be sensitive to the choice of $K$. In this study, we optimized the number of neighbors $K$ (in a range [1, 100]) with 1-fold and 5-fold cross-validation as SVM.

\subsubsection{Gaussian naive Bayes}
Gaussian naive Bayes (GNB) is a simple probabilistic classifier based on applying Bayes' theorem with strong (naive) independence assumptions between the features. It is a simple and effective algorithm that is easy to implement and works well on a variety of datasets. The algorithm works by using the training data to estimate the probability of each class, as well as the probability of each feature given a class. When given a new data point, the algorithm uses these probabilities to predict the class that is most likely to be associated with the data point. As in previous classifiers, we optimized the main hyperparameter in GNB, that is \verb|var_smoothing| (in a range [$10^{-1}, 10^{-100}$]) with 1-fold and 5-fold cross-validation.

The methods described above (feature extraction and the classifiers) are implemented in Python (tested on Python 3.7.4) using scikit-learn (\cite{scikit-learn}), Numpy \cite{Harris2020} and Pandas libraries with simple procedural implementations (no object-oriented programming and avoid a large loop). The source code is available at \url{https://github.com/bagustris/VBL-VA001}.

\section{Results and Discussion}\label{sec2}
We split the results and their discussions into two parts: extracted features and classification results. The former aims to show the differences among nine features at different axis vibration data for each machine condition. The latter shows the overall accuracy results of the three classifiers.

\subsection{Distinct Feature on Different Machine Condition}
\begin{figure}[htbp]
    \centering
    \includegraphics[width=\textwidth]{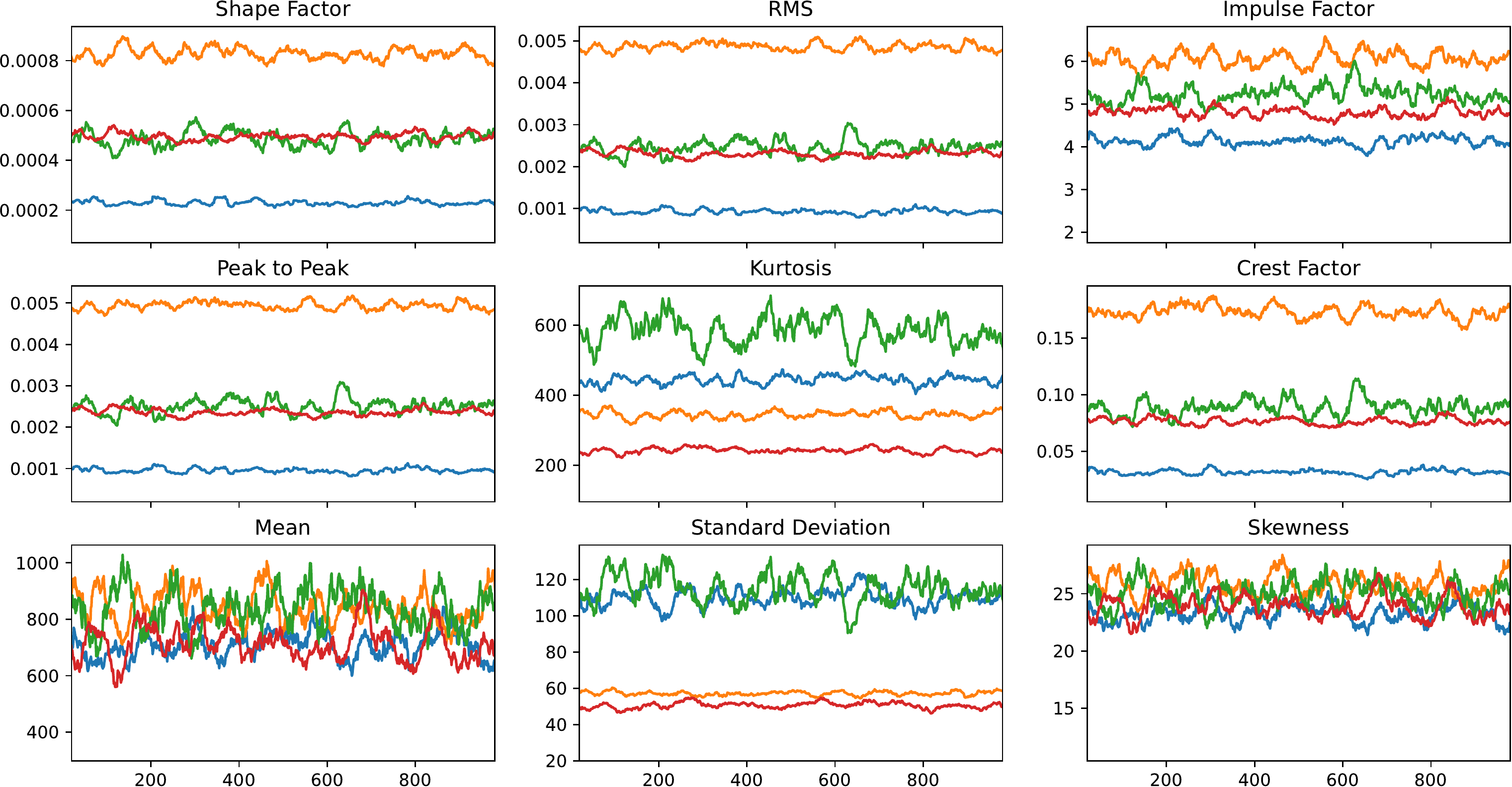}
    \caption{Plots of different feature values on the \textbf{x-axis} and the corresponding machine condition; blue: normal, orange: misalignment, green: unbalance, red: bearing fault.}
    \label{fig:feature-x}
\end{figure}

\begin{figure}[htbp]
    \centering
    \includegraphics[width=\textwidth]{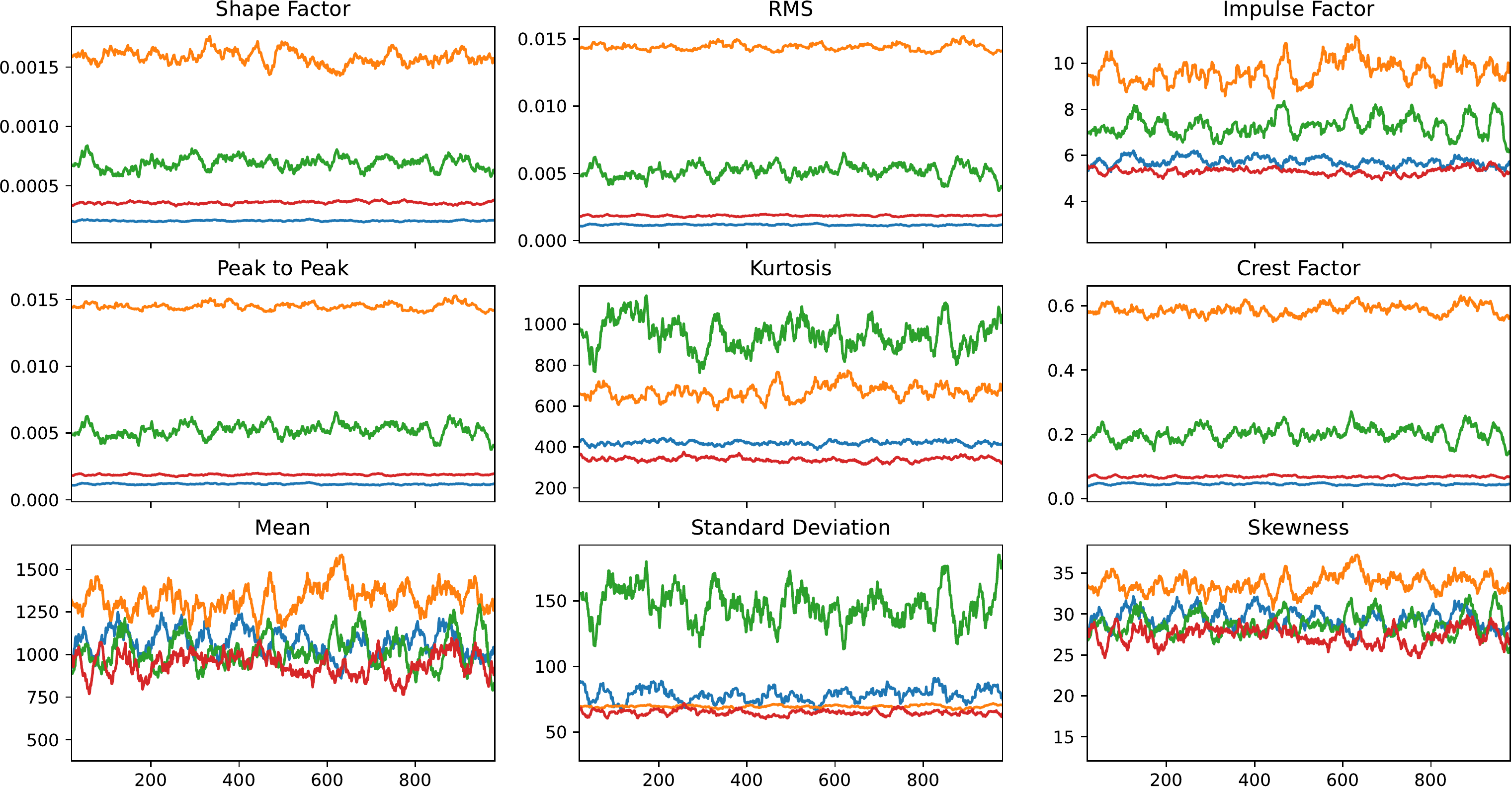}
    \caption{Plots of different feature values on the \textbf{y-axis} and the corresponding machine condition; blue: normal, orange: misalignment, green: unbalance, red: bearing fault.}
    \label{fig:feature-y}
\end{figure}

\begin{figure}[htbp]
    \centering
    \includegraphics[width=\textwidth]{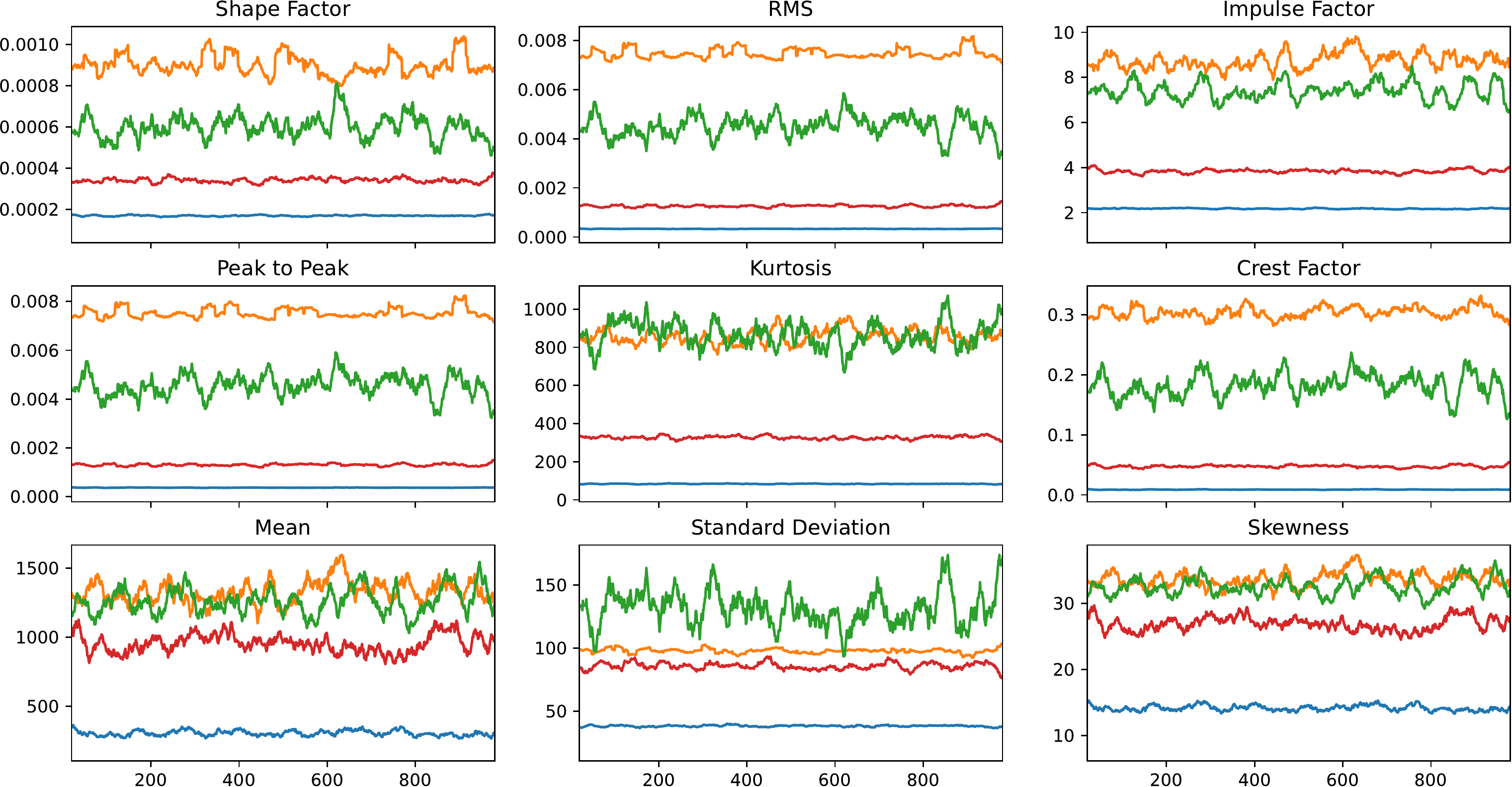}
    \caption{Plots of different feature values on the \textbf{z-axis} and the corresponding machine condition; blue: normal, orange: misalignment, green: unbalance, red: bearing fault.}
    \label{fig:feature-z}
\end{figure}

Figs. \ref{fig:feature-x}, \ref{fig:feature-y}, \ref{fig:feature-z} show the plot of nine feature values for different machine conditions on x, y, and z axes, respectively. It can be shown that, in general, our proposed features can discriminate a different condition of each machine. The most distinct features were observed in the y-axis where shape factor, RMS,  impulse factor, peak-to-peak, kurtosis, crest factor, and standard deviation of each machine condition are separable. Only mean and skewness features are confused among the machine conditions. By this observation, we can expect that the machine learning methods will be able to classify the machine condition with high accuracy.

Since the size of features is small, we can include all features from all axes. Also, the visualization of feature values among machine conditions shows distinct characteristics of each machine condition. Therefore, we also assumed no need to perform feature selection. However, all of those assumptions need verification by machine learning methods.

\subsection{Classification Results}
\begin{figure}[htbp]
    \centering
    \includegraphics[width=\textwidth]{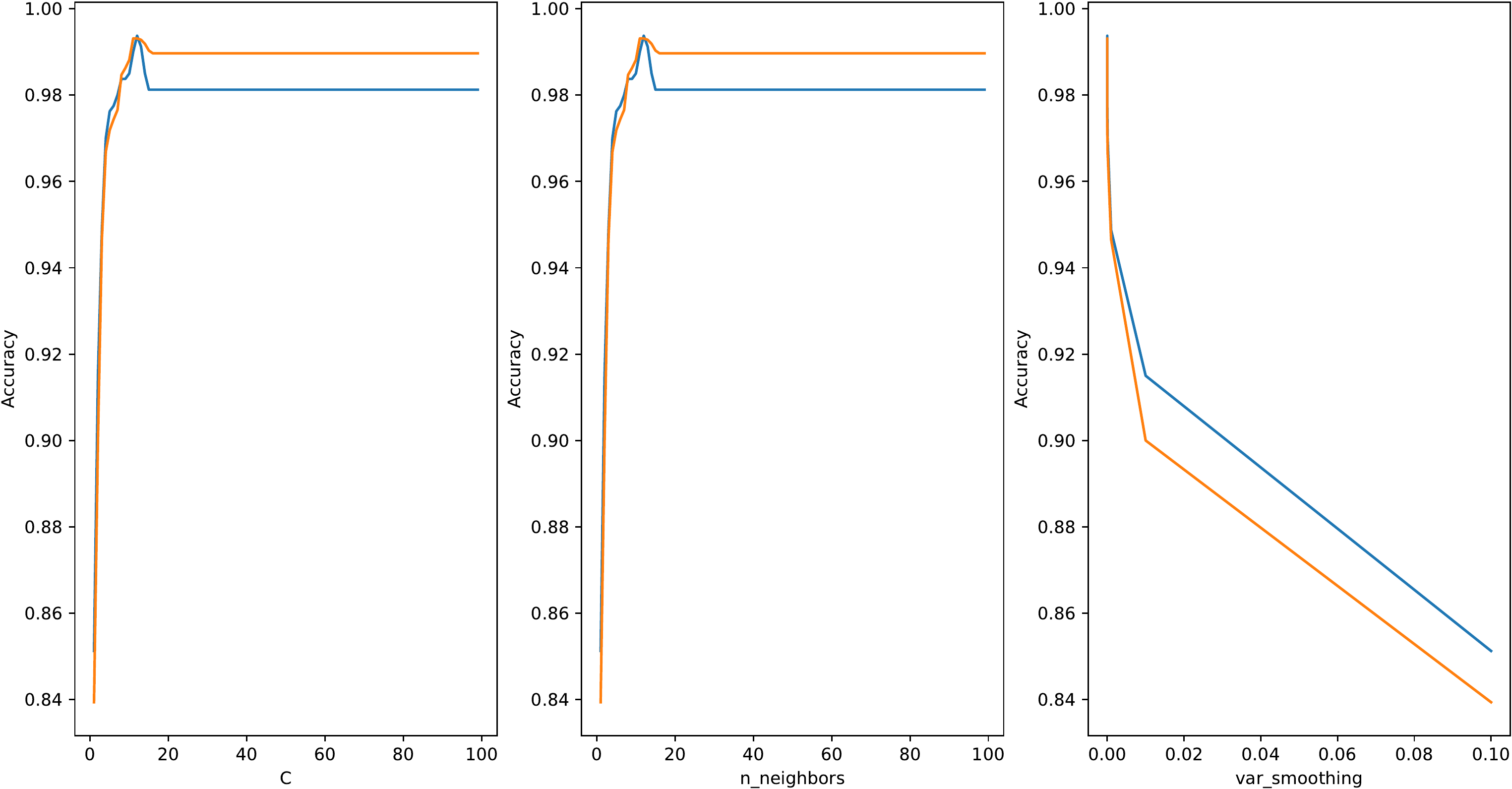}
    \caption{Parameter optimization results of SVM, KNN, and GNB; orange: training data, blue: test data. For 1-fold (shown above), the optimal parameters are 69 (C for SVM), 5 (K for KNN), and $10^{-11}$ (var\_smoothing for GNB). }
    \label{fig:parameter}
\end{figure}

\begin{figure}[htbp]
    \centering
    \includegraphics[width=\textwidth]{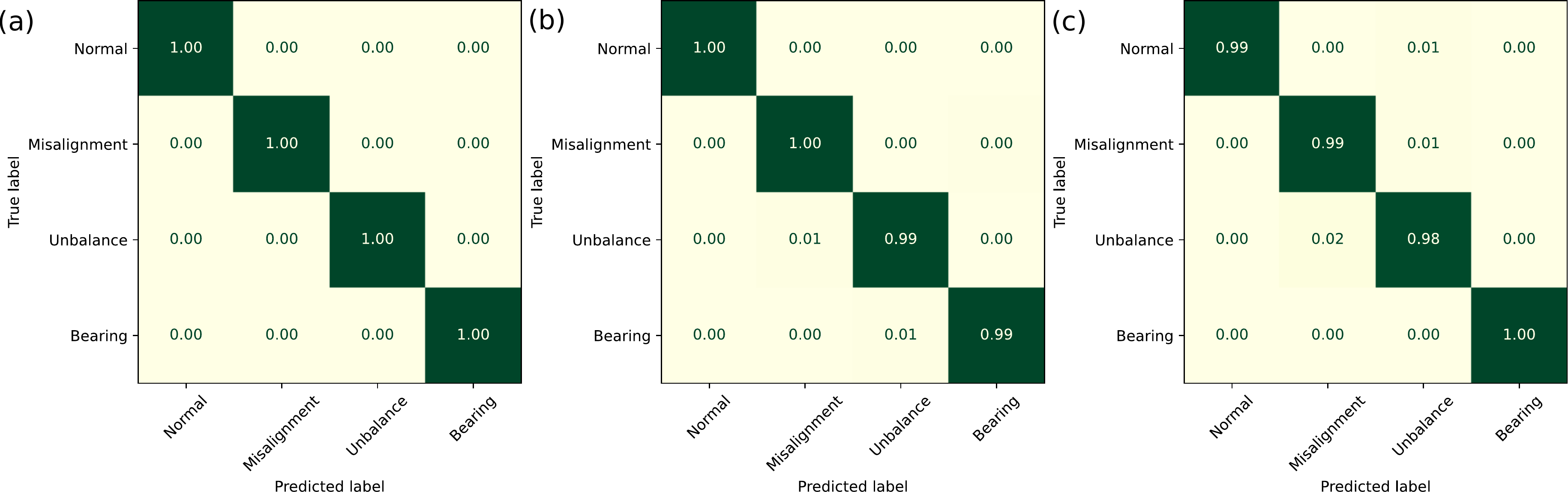}
    \caption{Confusion matrix on 1-fold test data for SVM (a), KNN (b), and GNB (c)}
    \label{fig:cm}
\end{figure}

The proposed balanced dataset allows us to evaluate the machine learning methods with a single metric, weighted accuracy. In the first step, we optimize the main parameter for each classifier (Table \ref{fig:parameter}), report accuracy results on 1-fold test and 5-fold cross-validation (Table \ref{tab:acc}), and show the confusion matrix on 1-fold test data (Fig. \ref{fig:cm}).

% parameter, shows for 5-fold also
Table \ref{tab:acc} shows the accuracy of 1-fold test and 5-fold cross-validation. For 1-fold, we obtained the best parameters are $C=69, K=5$, and \verb|var_smoothing| = 11, for SVM, KNN, and GNB, respectively (Table 9). For 5-fold, the best parameters are $C=93, K=1$, and \verb|var_smoothing| = 13. The best parameter for 5-fold is achieved by determining the maximum average of 5-fold tests on the given parameter search range.

Fig. \ref{fig:cm} shows the confusion matrix for 1-fold test. Perfect accuracy was obtained by the SVM method in this single-set test. In all classes, the recall was 100\% (shown as 1.00). The accuracies obtained by SVM in 5-fold are [99.875\%, 99.635\%, 99.5\%, 99.875\%, 100\%]. The worst case is 4 of 800 test data are incorrectly predicted by SVM (accuracy of 99.5\%). In a real implementation, the number of data for measurements can be added to mitigate this error. For instance, a single sample only contains five-second vibration data. More data for the same machine can be collected to increase the confidence level of the classifier. A recall of 0.01 corresponds to 2 samples, and a recall of 0.02 corresponds to 4 (from 200) samples in each class (for KNN and GNB).

% computation time
The computation time for the classification process is also short for SVM, KNN, and GNB. Given the input features in a CSV file (not the original vibration data), it only takes a minute to obtain the model's accuracy. The only process that takes longer time is the extraction process which takes about 10 minutes. The computation process was done on a PC with Intel Core i9-10850K CPU and 64GB RAM.

% create table two column and two row
\begin{table}[htbp]
    \centering
    \caption{Overall accuracy (\%) on 1-fold and 5-fold test data for SVM, KNN, and GNB.}
    \begin{tabular}{l c c}
        \hline
        Clasifier   & 1-fold	& 5-fold \\
        \hline
        SVM	        & 100	    & 99.75 \\
        KNN	        & 99.625	& 99.525 \\
        GNB	        & 99.5	    & 99.375 \\
        \hline
    \end{tabular}
    \label{tab:acc}
\end{table}

\section{Conclusion}\label{sec13}
In this paper, we present a new dataset for vibration analysis (machine condition classification) recorded from electric pumps in a laboratory environment. The dataset is balanced in four classes of machine conditions and contains 1000 samples for each condition. The evaluated conditions are normal, unbalance, misalignment, and bearing fault. We provided three classifiers as a baseline: SVM, KNN, and GNB. The inputs for the classifiers are nine statistical functions derived from the spectrum of vibration signals which show the distinct pattern for each machine condition. The results show that SVM has the best performance in this dataset which achieve overall accuracy of 99.75\% in 5-fold cross-validation. The high accuracy obtained by the SVM shows the potential use of the proposed dataset for machine learning research. Future research can be explored to obtain perfect accuracy and improve the classifiers' robustness and generalization, perhaps beyond the lab-scale environment.

\section*{Acknowledgments}
The authors would like to thank enDAQ for providing calibrated vibration sensor, LOG-0002-100G-DC-8GB-PC Shock and Vibration Sensor, and data acquisition system (enDAQ LAB) used in this research.
%%===========================================================================================%%
%% If you are submitting to one of the Nature Portfolio journals, using the eJP submission   %%
%% system, please include the references within the manuscript file itself. You may do this  %%
%% by copying the reference list from your .bbl file, paste it into the main manuscript .tex %%
%% file, and delete the associated \verb+\bibliography+ commands.                            %%
%%===========================================================================================%%

\bibliography{vibration-monitoring}% common bib file
%% if required, the content of .bbl file can be included here once bbl is generated
%%\input sn-article.bbl

%% Default %%
%%\input sn-sample-bib.tex%

\end{document}